%% file: main.tex
\input{release}
\ifdefined\isrelease
  \documentclass[sigconf]{acmart}
\else
  \documentclass[sigconf,authordraft]{acmart}
\fi

\usepackage{pgfplots}
\pgfplotsset{compat=1.18}
\usepackage{float}

\usepackage{wrapfig}

\ifdefined\isrelease
\else
  \usepackage[paperheight=11in, paperwidth=10in, margin=0.5in, marginparwidth=3.5in, marginparsep=0.4in]{geometry}
  \usepackage{mparhack}      
  \usepackage{marginnote}    
\fi

\ifdefined\isrelease
  \newcommand{\reviewnote}[1]{}
\else
  
  \newcommand{\reviewnote}[1]{\marginnote{\raggedright\footnotesize\textit{#1}}}
\fi

\ifdefined\isrelease
\else
\fi

\AtBeginDocument{%
  }

\ifdefined\isrelease
  \acmConference[Preprint Draft]{Preprint Draft}{2026}{}
  \acmBooktitle{Preprint Draft}
\else
  \setcopyright{acmlicensed}
  \copyrightyear{2026}
  \acmYear{2026}
  \acmDOI{XXXXXXX.XXXXXXX}
  \acmConference[Conf]{Conference}{2026}{San Jose}
  \acmISBN{978-1-4503-XXXX-X/2018/06}
\fi

\begin{document}

\title{If You Want Coherence, Orchestrate a Team of Rivals: Multi-Agent Models of Organizational Intelligence}

\ifdefined\skipauthors
  \author{Isotopes AI}
  \affiliation{%
    \country{USA}}
  \email{team@isotopes.ai}
\else
  \input{authors}
\fi

\begin{abstract}
AI Agents can perform complex operations at great speed, but just like all the humans we have ever hired, their intelligence remains fallible. Miscommunications aren't noticed, systemic biases have no counter-action, and inner monologues are rarely written down. 

We did not come to fire them for their mistakes, but to hire them and provide a safe productive working environment. We posit that we can reuse a common corporate organizational structure: teams of independent AI agents with strict role boundaries can work with common goals, but opposing incentives. Multiple models serving as a team of rivals can catch and minimize errors within the final product at a small cost to the velocity of actions. In this paper we demonstrate that we can achieve reliability without acquiring perfect components, but through careful orchestration of imperfect ones.

This paper describes the architecture of such a system in practice: specialized agent teams (planners, executors, critics, experts), organized into an organization with clear goals, coordinated through a remote code executor that keeps data transformations and tool invocations separate from reasoning models. Rather than agents directly calling tools and ingesting full responses, they write code that executes remotely; only relevant summaries return to agent context. By preventing raw data and tool outputs from contaminating context windows, the system maintains clean separation between perception (brains that plan and reason) and execution (hands that perform heavy data transformations and API calls). We demonstrate the approach achieves over 90\% internal error interception prior to user exposure while maintaining acceptable latency tradeoffs. A survey from our traces shows that we only trade off cost and latency to achieve correctness and incrementally expand capabilities without impacting existing ones.
\end{abstract}



\maketitle


\input{sections/introduction}

\input{sections/related-work}

\input{sections/architecture}

\input{sections/evaluation}

\input{sections/discussion}

\input{sections/conclusion}

\ifdefined\isrelease
\else
\begin{acks}
[TODO: Add acknowledgments]
\end{acks}
\fi

\bibliographystyle{ACM-Reference-Format}
\bibliography{main}

\end{document}

%% file: release.tex
\def\isrelease{1}

\PassOptionsToClass{nonacm}{acmart}

%% file: authors.tex
\author{Gopal Vijayaraghavan}
\affiliation{%
  \institution{Isotopes AI}
  \country{USA}}
\email{gopal@isotopes.ai}

\author{Prasanth Jayachandran}
\affiliation{%
  \institution{Isotopes AI}
  \country{USA}}
\email{prasanth@isotopes.ai}

\author{Arun Murthy}
\affiliation{%
  \institution{Isotopes AI}
  \country{USA}}
\email{arun@isotopes.ai}

\author{Sunil Govindan}
\affiliation{%
  \institution{Isotopes AI}
  \country{USA}}
\email{sunil@isotopes.ai}

\author{Vivek Subramanian}
\affiliation{%
  \institution{Isotopes AI}
  \country{USA}}
\email{vivek@isotopes.ai}

\renewcommand{\shortauthors}{Vijayaraghavan et al.}

%% file: sections/introduction.tex
\section{Introduction}
\label{sec:intro}

Large language models have demonstrated remarkable capabilities across diverse tasks, yet their deployment in production systems remains hampered by a fundamental challenge: unreliability.\reviewnote{Strong opening. Consider citing recent production failures (e.g., chatbot hallucinations in customer service) to ground the problem.} When a single LLM hallucinates, misinterprets context, or makes a logical error, there is no mechanism to catch the mistake before it propagates downstream. In high-stakes domains, such as financial analysis, healthcare decision support, and legal document processing, a single uncaught error can have severe consequences.

The standard approach to deploying LLM-based systems mirrors hiring a single brilliant analyst: you craft an elaborate prompt, invoke one model, and trust the output. This works well for demonstrations and low-stakes applications. But just as no responsible organization would rely on a single employee to handle critical operations, verify their own work, and catch their own mistakes, we should not architect production AI systems around single-agent execution, even with self-review.

\subsection{From Single AI Person to AI Office}

Consider the difference between a solo worker and an entire office. A talented individual working alone has inherent limitations: systematic biases go unchecked, difficult edge cases get missed, and when mistakes occur, there is no safety net. An office, in contrast, achieves reliability through organizational structure: specialization allows experts to focus on their strengths, redundancy catches errors through peer review, oversight provides hierarchical validation, and collaborative problem-solving distributes complex tasks across appropriate specialists.

Just as no responsible hiring manager would recruit exclusively from a single institution, we should not architect AI systems around models from a single provider. Different model families exhibit complementary strengths and failure modes. Relying on a single provider creates a monoculture: cost structures become inflexible, capabilities are bounded by that provider's architecture, and performance cannot be optimized for diverse task requirements. A team drawing from multiple model providers achieves cognitive diversity that no single source can provide, matching specialized models to appropriate tasks while reducing systematic biases and expanding the solution space for complex problems.

Traditional single-agent LLM systems exhibit the qualities of the solo worker: simplicity, speed and lower cost. However, as tasks require more tools and specific training to use them, the limitations of a jack of all trades are exposed, with too many tools to switch between creating context switching overheads.

This paper presents an alternative architecture that creates an AI office full of specialists rather than relying on a single generalist.\reviewnote{Good framing. The office metaphor is intuitive---make sure this carries through consistently to Section 3.}

\subsection{Achieving Coherence}

Central to this work is the concept of \textbf{coherence}: the quality of forming a unified whole. Applied to large organizations, coherence is maintained by opposing forces holding outputs within an acceptable zone. Each stakeholder pushes in a different direction: one for completeness, another for practicality, a third for correctness. Their conflicting incentives create boundaries that prevent drift. Advancement requires consensus among stakeholders with veto authority, forcing results into the intersection where all opposing forces find the output acceptable. This is not compromise or majority voting; it is the discipline of satisfying rivals simultaneously.

We apply this principle to multi-agent AI systems. The planner pushes for clarity and completeness, the executor pushes for pragmatic implementation, and the critic pushes for correctness and standards compliance. A single-agent system lacks these opposing forces: the same entity that craves completion also evaluates completion, with no counterbalance. Our AI office achieves coherence when agents with conflicting roles reach consensus that the output is acceptable.

Our multi-agent system achieves reliability through organizational principles: specialized agents occupy distinct roles (planner, executor, critic), hierarchical veto authority prevents errors from propagating, and pre-declared acceptance criteria establish clear quality gates.\reviewnote{Clear contribution. Consider adding upfront: "We demonstrate 92.1\% error recovery on 522 production sessions" to establish scale immediately.} Just as financial reports require independent audits rather than accountant self-certification, our architecture ensures that code writers cannot declare their own work complete; only independent critics with veto authority can approve outputs for advancement.

\subsection{Theoretical Foundation}

Our approach builds on two foundational concepts. First, Reason's Swiss cheese model~\cite{Reason1990} describes how multiple imperfect layers of defense can achieve system reliability: even if each layer has holes (failure modes), if we can ensure the holes are misaligned, hazards cannot propagate through all layers simultaneously. We apply this to AI systems through team-based validation, where multiple agents with different failure modes create defense-in-depth.

Second, we treat miscommunication as the fundamental problem instead of treating it as an avoidable error. Reliable communication over a noisy channel can operate reliably, but at a much lower bandwidth than the maximum possible. Shannon's channel capacity theorem~\cite{Shannon1948} showed that reliable communication is achievable even when individual transmissions are noisy, establishing a precise theoretical boundary that you cannot exceed. The theorem introduced mutual information as the key quantity linking channel input and output, measuring how much information successfully transfers despite noise.\reviewnote{Clever analogy, but needs clarification: how does "mutual information" specifically map to inter-agent communication? What is being transmitted/received in your model?} We apply this insight to multi-agent systems by treating inter-agent communication as a noisy channel. We expand our system by encoding data moving across agents with verbosity and retries as redundancy, achieving reliable information flow between unreliable model components. This idea of redundancy for safety mirrors "Shisa Kanko," the Japanese railway practice of pointing and calling to prevent errors through deliberate repetition. This overhead reduces throughput, which we quantify later in the paper.

\ifdefined\isrelease
\else
\fi

The goal is not to eliminate component unreliability, an impossible task, but to achieve system reliability that exceeds what any individual component could provide.

\subsection{Architecture Overview}

Our system comprises 50+ specialized agents organized into teams with distinct roles: planners generate execution strategies, executors perform work, critics validate outputs against pre-declared criteria, and a remote code executor maintains clean separation between reasoning and data transformation. Critical to reliability is hierarchical veto authority: critics can reject outputs entirely, triggering team-internal retry without re-planning. This prevents the accountability problems with council based voting, where a specialized critic with domain expertise can halt propagation even when consensus finds nothing wrong.

The remote code executor~\cite{Mei2024AIOS} keeps data transformations and tool invocations separate from reasoning models, preventing raw data and tool outputs from contaminating agent context windows. Rather than agents directly calling tools (which would inject full responses into their context), they write code that invokes tools like MCP servers; tool responses remain in the remote execution layer, with only relevant summaries returning to agents. Agents perceive results through summaries and schemas rather than full datasets or raw API responses, maintaining the separation between perception (brains that plan and reason) and execution (hands that perform heavy data transformations and API calls). This architecture solves the tool complexity trap: rather than burdening a single agent with 50+ tools and an unwieldy prompt containing massive datasets and tool outputs, each specialized agent reasons about what to do while the remote executor handles the dirty work.

\subsection{Contributions}

This work makes the following contributions:

\begin{itemize}
  \item \textbf{Organizational reliability for AI systems}: Demonstration that standard organizational practices for large engineering teams transfer effectively to multi-agent AI systems.
  \item \textbf{Context Ray Tracing}: A message visibility mechanism that controls information flow between agents at different hierarchical levels. Rather than broadcasting all information to all agents, representatives attend cross-team coordination points and relay only relevant summaries, analogous to organizational meetings where delegates discuss and decide on behalf of their teams.
  \item \textbf{Data Isolation}: Separation of reasoning from execution by keeping data transformations and tool invocations in a remote layer, preventing context contamination with raw data. The agents can request multiple summaries, samples and outliers instead, so the working set size can be much larger than the context window.
  \item \textbf{Multi User Interactivity}: Integration of multiple human stakeholders into the organizational structure, enabling approvals, reviews, and escalations to be distributed across different users rather than funneling all interactions through a single operator.
  \item \textbf{Post-hoc Audits}: A bidirectional traversal mechanism for the action graph that supports multiple audit workflows: sidebar questions during execution, backward tracing from results to debug errors, citation maintenance linking outputs to source evidence, and exposure analysis to identify downstream impacts when cited facts are corrected.
\end{itemize}

The remainder of this paper is organized as follows: Section~\ref{sec:related} surveys related work on multi-agent systems and ensemble methods; Section~\ref{sec:architecture} details the system architecture; Section~\ref{sec:evaluation} presents evaluation results; Section~\ref{sec:discussion} discusses implications and limitations; Section~\ref{sec:conclusion} concludes.

%% file: sections/related-work.tex
\section{Related Work}
\label{sec:related}

Recent research on LLM-based systems has studied (i) failure modes and reliability challenges in multi-agent LLM systems, (ii) trust, risk, and security management for agentic settings, (iii) LLM ensembles and consensus-based validation for improving reliability, (iv) multi-LLM collaboration in high-stakes domains, and (v) evaluation and benchmarking for agent safety and enterprise deployment requirements.

\subsection{Multi-Agent LLM Systems}

Cemri et al.~\cite{Cemri2025} introduce MAST-Data, a dataset of over 1,600 annotated traces collected across 7 multi-agent system frameworks. They build the MAST failure taxonomy through analysis of 150 traces guided by expert human annotators ($\kappa = 0.88$), identifying 14 failure modes grouped into three categories: system design issues, inter-agent misalignment, and task verification failures.

Huang et al.~\cite{Huang2024Resilience} investigate multi-agent collaboration in the presence of faulty agents (described as ``clumsy'' or malicious). They propose AutoTransform and AutoInject to introduce mistakes into agents' responses, study the resilience of multiple system structures, and report that a hierarchical structure yields the lowest performance drop (5.5\%) compared to 10.5\% and 23.7\% for two other structures. They also introduce resilience mechanisms named Challenger (agents challenge others' outputs) and Inspector (an additional reviewing agent), reporting recovery of up to 96.4\% of errors made by faulty agents.

Raza et al.~\cite{Raza2025TRiSM} adapt and extend the TRiSM (Trust, Risk, and Security Management) framework for LLM-based agentic multi-agent systems, describing a framework structured around pillars including Explainability, ModelOps, Security, and Privacy/Governance. They propose a risk taxonomy spanning issues from coordination failures to prompt-based adversarial manipulation, and introduce two evaluation metrics: Component Synergy Score (CSS) and Tool Utilization Efficacy (TUE).

Wang et al.~\cite{Wang2024MultiPersona} propose Solo Performance Prompting (SPP), a prompting method that enables a single LLM to act as a multi-persona self-collaborator by dynamically identifying and simulating personas based on the task input. They report that assigning multiple fine-grained personas improves problem-solving compared to using a single or fixed number of personas, reduces factual hallucination while maintaining reasoning ability, and observe that the ``cognitive synergy'' phenomenon appears in GPT-4 but not in less capable models.

\subsection{LLM Ensemble and Consensus Methods}

Naik~\cite{Naik2024} proposes a framework that repurposes ensemble methods for content validation through model consensus, motivated by reliability needs in high-stakes settings. The paper reports tests across 78 complex cases requiring factual accuracy and causal consistency, with precision improvements from 73.1\% to 93.9\% (two models) and 95.6\% (three models), and reports inter-model agreement $\kappa > 0.76$ while retaining sufficient independence to catch errors through disagreement. The work also notes constraints including multiple-choice requirements and processing latency.

Kamen and Kamen~\cite{Kamen2025} propose an ensemble large language model framework (eLLM) for unstructured text categorization, describing how aggregation can address issues including inconsistency, hallucination, category inflation, and misclassification. They report up to 65\% F1-score improvement over the strongest single model, formalize aggregation criteria via a collective decision-making model, and evaluate ten LLMs under zero-shot conditions on a human-annotated corpus of 8,660 samples using the IAB hierarchical taxonomy.

Chen et al.~\cite{Chen2025LLMEnsemble} provide a survey of LLM ensemble research, describing a taxonomy that classifies methods into ensemble-before-inference, ensemble-during-inference, and ensemble-after-inference categories. They also cover benchmarks and applications and provide a curated list of relevant work.

\subsection{Multi-Model Collaboration in High-Stakes Domains}

Sanchez et al.~\cite{Sanchez2025} apply multi-LLM collaboration to medication recommendation, building on prior work they describe as LLM Chemistry (a measure of collaborative compatibility among LLMs). They describe a two-stage collaboration mechanism and explicitly state that they extend the evaluation step with a consensus step that transforms diverse (sometimes conflicting) model outputs into a unified decision. They evaluate the approach on medication recommendation and report improved accuracy and stability over other ensemble baselines.

\subsection{Evaluation and Benchmarking of Agent Safety}

Zhang et al.~\cite{Zhang2024AgentSafetyBench} introduce Agent-SafetyBench, a benchmark with 349 interaction environments and 2,000 test cases covering 8 categories of safety risks and 10 common failure modes. They report evaluating 16 popular LLM agents and finding that none achieves a safety score above 60\%. They also identify two safety defects: lack of robustness and lack of risk awareness, and report that defense prompts alone may be insufficient.

Mohammadi et al.~\cite{Mohammadi2025} survey evaluation and benchmarking of LLM agents and propose a two-dimensional taxonomy organizing work by (i) evaluation objectives (including behavior, capabilities, reliability, safety) and (ii) evaluation process (including interaction mode, datasets/benchmarks, metric computation methods, tooling). They highlight enterprise-specific challenges, including role-based access control and the need for reliability guarantees for audit and compliance purposes, alongside long-horizon interaction and compliance requirements.

\subsection{Positioning and Contributions}

While prior work establishes the potential of multi-agent systems and ensemble methods, our contribution lies in the integration and operationalization of these concepts into a production-ready architecture. Prior work is largely theoretical or evaluated at small scale; we evaluate on 522 production sessions with quantified cost-benefit tradeoffs (38.6\% overhead). We combine:

\begin{itemize}
  \item \textbf{Role-based specialization} with strict boundaries preventing context contamination
  \item \textbf{Hierarchical veto authority} (not consensus voting) inspired by organizational structures
  \item \textbf{Pre-declared acceptance criteria} (not emergent verification) following test-driven development principles
  \item \textbf{Message passing kernel abstraction} to maintain legibility across actions
  \item \textbf{Swiss cheese layered validation} where multiple imperfect checkers with misaligned failure modes catch errors
\end{itemize}

Unlike probabilistic consensus approaches~\cite{Naik2024}, our critics have absolute veto authority. Unlike democratic voting schemes~\cite{Kamen2025}, our architecture is hierarchical with asymmetric power. Unlike general ensemble methods~\cite{Chen2025LLMEnsemble}, our agents occupy specialized roles rather than serving as interchangeable models.

These technical mechanisms combine to deliver the five contributions outlined in Section~\ref{sec:intro}: (1) \textbf{Organizational reliability} emerges from hierarchical veto authority and pre-declared acceptance criteria transferring software engineering practices to AI; (2) \textbf{Context Ray Tracing} is enabled by message passing kernel abstraction and role-based specialization creating selective visibility; (3) \textbf{Data Isolation} prevents context contamination through strict role boundaries and remote execution; (4) \textbf{Multi User Interactivity} leverages the organizational structure for human approvals and escalations; (5) \textbf{Post-hoc Audits} are possible because message passing provides complete traversal of all agent interactions. This integration creates a practical framework for deploying LLM-based systems in production environments where correctness is non-negotiable.

%% file: sections/architecture.tex
\section{System Architecture}
\label{sec:architecture}








\subsection{Multi-Agent Architectures: From Delegation to Organizational Oversight}
\label{sec:multi-agent-comparison}

The term ``multi-agent'' obscures more than it reveals. A system with two agents and a system with fifty agents organized into reviewing committees are both ``multi-agent,'' yet they differ in a property that matters far more than agent count: \textit{where errors get caught}. We distinguish three architectural patterns by this criterion.

\subsubsection{Tool Chaining: Sequential Execution, No Oversight}

The simplest multi-agent pattern chains tools or functions sequentially. An agent reasons, calls a tool, incorporates the result, reasons again, calls another tool. Modern agentic systems (from ReAct~\cite{Yao2023ReAct} to function-calling assistants) follow this pattern. The agent may invoke dozens of tools, but authority never leaves the single reasoning loop.

Tool chaining provides flexibility and simplicity. But errors in early steps compound through later ones.\reviewnote{Evidence needed: cite or show error propagation rates from tool chaining systems. How much worse is cascading error?} The agent that made a mistake is the same agent evaluating whether a mistake was made. Self-review is better than no review, but it shares the blind spots of the original reasoning.

\subsubsection{Sub-Agent Parallelization: Fan-Out for Throughput}

Sub-agent architectures extend tool chaining with parallelism. A parent agent decomposes work, spawns child agents to execute subtasks concurrently, and aggregates their outputs. Anthropic's research system~\cite{Anthropic2025} exemplifies this pattern, deploying 10+ sub-agents for complex investigations with clearly divided responsibilities.

Sub-agents dramatically improve throughput for independent tasks. Search ten databases simultaneously; fetch from multiple APIs in parallel; explore alternative hypotheses concurrently. When subtasks are genuinely independent, sub-agents provide clean parallelization.

The limitation emerges when subtasks are not independent. Sub-agents cannot see sibling work. When one sub-agent assumes fiscal year ends in December and another assumes March, the parent receives conflicting outputs with no visibility into the conflicting assumptions. Cognition's analysis~\cite{Cognition2025} identifies this as the primary failure mode of naive multi-agent systems: ``actions carry implicit decisions, and conflicting decisions carry bad results.'' The parent aggregates outputs but cannot detect conflicts it cannot see.

\subsubsection{Organizational Council: Stage-Gated Oversight}

Our architecture takes a fundamentally different approach. Rather than optimizing for throughput, we optimize for a different property: \textit{errors should die in committee, not surface to users}.

The organizational council interposes critics at multiple stages of execution. Work does not flow directly from producer to user. It flows from producer to critic, and only approved work advances. Rejected work triggers internal retry; the user never sees the first draft with the wrong join logic, the chart that misrepresented the trend, or the analysis that violated accounting standards.\reviewnote{Concrete examples strengthen the argument. Add 1-2 specific examples of errors caught at each layer from your traces.}

Multiple specialized critics (for plans, code, outputs, and domain methodology) each serve as filters with distinct failure modes. The Swiss cheese principle (Section~\ref{sec:intro}) applies: errors that slip through one filter encounter another.\reviewnote{Key assumption: failure modes are "misaligned". Verify empirically---show correlation matrix of error types caught by each critic layer.} Critics hold veto authority, not advisory input; the mechanics of this authority structure are detailed in Section~\ref{sec:team-execution}.

\subsubsection{The Architectural Tradeoff}

Table~\ref{tab:error-handling} summarizes how each pattern handles errors.

\begin{table}[htbp]
\centering
\caption{Error Handling Across Multi-Agent Architectures}
\label{tab:error-handling}
\small
\begin{tabular}{@{}p{2.1cm}p{2.5cm}p{2.8cm}@{}}
\toprule
\textbf{Pattern} & \textbf{Error Detection} & \textbf{User Exposure} \\
\midrule
Tool chaining & Self-review only & Errors surface directly \\
Sub-agent & Parent aggregates & Conflicts undetected \\
Org. council & Stage-gated critics & Only approved outputs \\
\bottomrule
\end{tabular}
\end{table}

The tradeoff is cost. Tool chaining is cheapest: one reasoning loop, minimal overhead. Sub-agents add parallelization cost but maintain single-point aggregation. Organizational councils multiply inference calls: every critic evaluation, every retry loop, every SME consultation adds latency and API cost.

We argue this cost is justified for high-stakes domains. A financial analysis that reaches the user with an incorrect calculation damages trust and may drive poor decisions. An analysis that takes longer but arrives correct builds confidence in the system. The organizational council trades latency for reliability, the same tradeoff human organizations make when they institute review processes, approval chains, and audit requirements.

\subsubsection{Combining Patterns}

These patterns compose rather than compete. Our architecture uses organizational council structure at the macro level (planners, executors, critics with stage-gated oversight) while individual teams may employ sub-agent parallelization internally. CodingInnerLoopTeam operates as a council (writer $\rightarrow$ executor $\rightarrow$ critic), but the executor may spawn sub-agents for parallel file operations.

The principle: \textbf{sub-agents for throughput within trust boundaries; organizational oversight across trust boundaries}. Parallel execution is safe when subtasks are genuinely independent and outputs will be validated before advancing. Stage-gated review is essential when errors in one component affect correctness in another, or when outputs will reach users.

The contract with users differs fundamentally from single-agent or sub-agent systems. Those systems return what was produced. Organizational councils return what was \textit{approved}: outputs that survived scrutiny from critics who did not produce them, evaluated against criteria declared before execution began.

\subsection{Architecture of the AI Office}
\label{sec:office-architecture}

\subsubsection{Agent Specialization and Roles}

The system comprises 50+ specialized agents organized by function and responsibility. Rather than treating all agents identically, each type optimizes for its specific role:

\begin{itemize}
  \item \textbf{Planners}: Parse user queries, refine for clarity, retrieve relevant context from memory and metadata, construct execution DAGs, enforce domain guardrails, and manage plan orchestration. These agents handle semantic understanding and intention modeling.

  \item \textbf{Executors}: Orchestrate execution of plans, route work to appropriate specialists, manage iterative refinement loops, and coordinate across writing, execution, and critique phases. Executors handle deterministic state management and handoff sequencing.

  \item \textbf{Data Writers}: Specialized per data source (SQL databases, spreadsheets, Python environments, APIs, etc.). Each writer generates appropriate code for its target system, enabling agents to work with data sources through unified abstractions while leveraging source-specific optimizations.

  \item \textbf{Critics}: Domain-specialized validators that operate at different abstraction levels. Code critics verify correctness and security; output critics validate against user intent and pre-declared acceptance criteria; plan critics verify soundness of execution DAGs. Each critic can veto independent of others.

  \item \textbf{Responders}: User-facing agents that handle approvals, escalations, clarification requests, and result synthesis. These agents manage the human-in-the-loop checkpoints and human-readable output generation.

  \item \textbf{Summarizers}: Distill intermediate results into compact summaries for downstream agents, enabling context minimization without losing decision-relevant information.

  \item \textbf{SME Experts}: Domain specialists in financial analysis, visualization, reconciliation, and other vertical domains. Provide guidance during planning for complex queries requiring domain knowledge.

  \item \textbf{Coordinators}: Manage execution orchestration, route requests to execution teams based on task classification, and synchronize handoffs between phases.
\end{itemize}

Agent selection and composition are determined dynamically by the planner based on each user request. Rather than using a fixed team structure, each prompt causes the planner to compose specialized teams containing writers, critics, executors, and summarizers selected based on the task requirements.\reviewnote{Interesting: show an example. How does the planner decide to spawn Python vs SQL vs Chart writers? Is this rule-based or learned? Current presentation is vague.} The planner analyzes the user request, identifies required data sources and domain expertise, and constructs a team roster tailored to the specific problem.

The execution sequence follows a fixed pattern: \textbf{Planner first, Critics last, with iterative loops in between}. The planner generates an execution DAG with pre-declared success criteria and acceptance gates. Writers and executors then perform work, producing outputs. Critics evaluate these outputs against the success criteria; if approved, results advance; if rejected, the internal team retries without replanning. This loop repeats until critics approve or escalation occurs.

Team composition also leverages vendor diversity: writers and critics run on different model providers to achieve cognitive diversity and avoid monoculture. A writer might generate code using one provider's model, while a critic from another provider validates the output, catching errors that shared training would miss. This cross-vendor strategy optimizes for different strengths: fast, economical writers for routine generation; more capable critics for rigorous validation.

\begin{figure*}[htbp]
\centering
\includegraphics[width=0.6\linewidth]{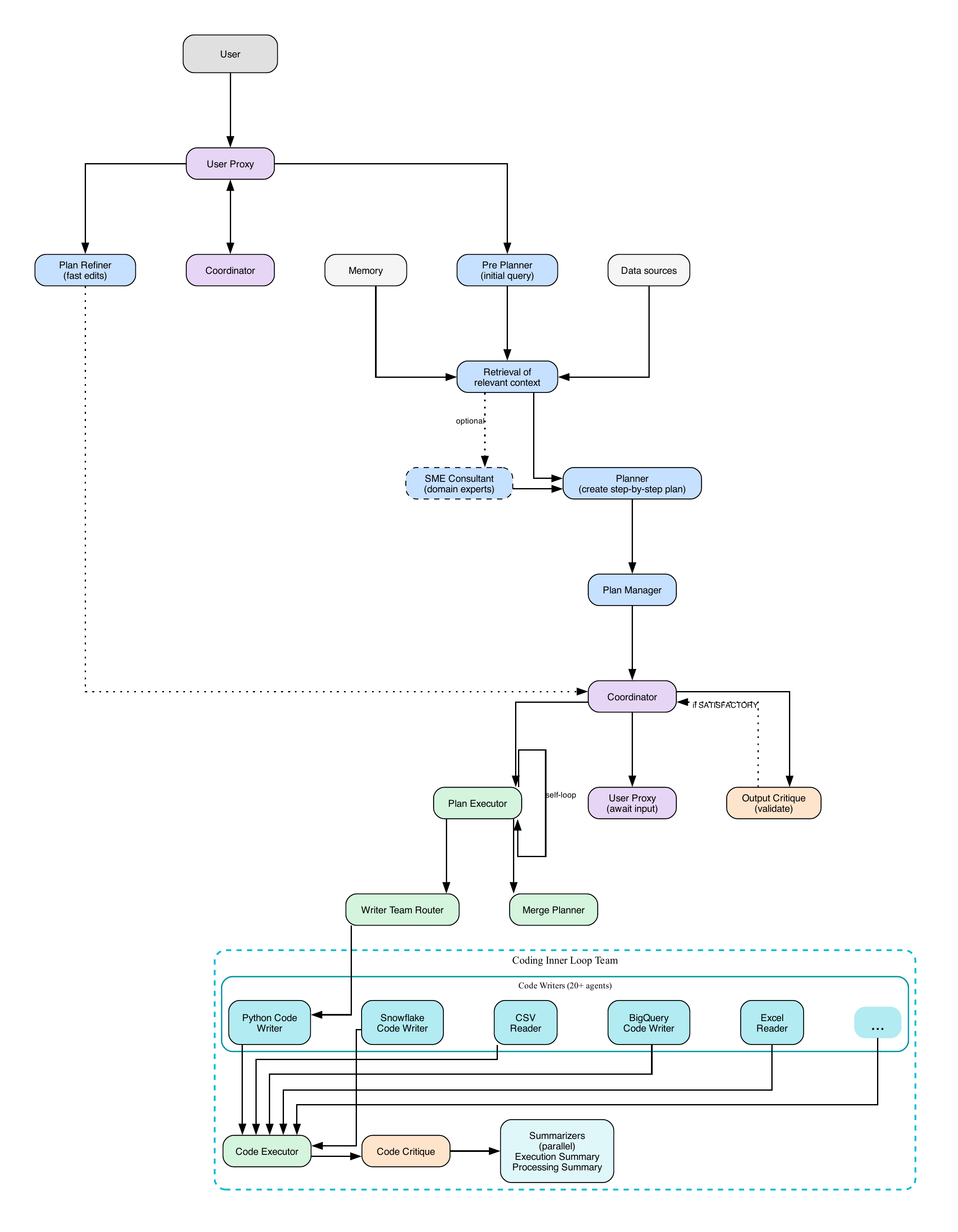}
\caption{Multi-agent FSM architecture with color-coded execution phases.}
\Description{A hierarchical directed graph showing the finite state machine architecture of a multi-agent AI system with spaced agent names. At the top, User connects to User Proxy gateway with a single arrow. User Proxy has bidirectional arrow to Coordinator and unidirectional arrows to Pre Planner (labeled "initial query") and Plan Refiner (labeled "fast edits"). On the left side, Data sources and Memory boxes feed into Retrieval of relevant context. The flow continues downward through planning agents (SME Consultant labeled "domain experts" as optional dotted path, Planner labeled "create step-by-step plan", Plan Manager), then to a second Coordinator instance. Plan Refiner has a dotted line to this bottom Coordinator. The bottom Coordinator splits to three nodes at the same level: Plan Executor, User Proxy (labeled "await input"), and Output Critique (labeled "validate"). Plan Executor has a self-loop edge and routes to Writer Team Router and Merge Planner. A dashed box labeled "Coding Inner Loop Team" contains the execution sequence: Writers cluster → Code Executor → Code Critique → Summarizers (parallel execution of Execution Summary and Processing Summary). Output Critique has one dotted feedback arrow to Coordinator labeled "if SATISFACTORY". All agent names use spaces for readability. Arrows use curved diagonal routing for natural flow.}
\label{fig:agent-architecture}
\end{figure*}

FSM-based routing orchestrates this dynamic composition. The system escalates to senior agents (larger, more capable models) when disagreement occurs between critics or when standard agents fail. See Section~\ref{sec:adaptive-allocation} for details on how model allocation scales capability with task complexity.
Figure~\ref{fig:agent-architecture} illustrates the complete multi-agent FSM architecture with these workflow stages integrated across planning, execution, and validation phases.

\subsubsection{The Data Infrastructure Layer}

Five-layer data kernel providing standardization and routing:

\begin{itemize}
  \item \textbf{Intake \& Standardization}: Connectors abstract 15+ data sources (Snowflake, BigQuery, PostgreSQL, Stripe, QuickBooks, Workday, CSV, Excel, JSON, PDF, OpenMetadata); normalized to common internal format
  \item \textbf{Transformation \& Routing}: DataWrangler provides 50+ composable primitives; agents request operations; kernel executes transformations; router directs outputs to appropriate next agent
  \item \textbf{Validation}: GuardrailsAgent enforces policies; column-level lineage tracking; schema validation; data quality checks before agent processing
  \item \textbf{Synthesis}: OutputCritique validates team outputs against acceptance criteria; approved results formatted for downstream consumption; rejected results trigger team re-execution
  \item \textbf{Persistence}: Vector memory (PostgreSQL + pgvector) captures insights; indexed by scope (account/user/session); semantic search across all prior analyses
\end{itemize}

\subsubsection{Agent Coordination and Communication}

Drawing from AIOS~\cite{Mei2024AIOS}, our system operates like a cooperative message passing kernel where agents yield control rather than running entirely concurrently, communicating through structured messages. The following mechanisms enable team-based orchestration with clear role boundaries:

\begin{itemize}
  \item \textbf{Message passing}: Agents communicate through Pydantic-validated structured messages, avoiding plain text tokens for inter-agent communication. Each message is strongly typed, enabling type safety across agent boundaries and eliminating parsing ambiguity. Message visibility is filtered based on FSM state, ensuring agents receive only messages relevant to their role and phase. This architecture is analogous to inter-process communication in operating systems, where structured protocols replace ad-hoc text formats.
  \item \textbf{Handoff pattern}: Agents delegate to specialists; PlanManager routes to PlanExecutor → CodingInnerLoopTeam → Critics; explicit role boundaries prevent cross-contamination. Critically, handoff abstracts internal execution details: if a task requires three retries to succeed, the manager layer receives only the final approved output, not the retry history or intermediate failures. Only semantic memory captures (e.g., insights about data patterns or edge cases discovered during retries) propagate upward to improve future attempts; execution mechanics remain encapsulated within team boundaries.
  \item \textbf{Checkpointing}: At decision points and after each phase, the entire ordered state of all agents (including their context, reasoning, intermediate results, and execution status) is completely serialized and persisted as a checkpoint. This enables users to keep sessions active indefinitely: a session can pause, and users can return days or weeks later to resume from the exact checkpoint. More powerfully, users can time-travel backward to any previous checkpoint to undo changes, explore alternative paths, or retry with different parameters, without losing work or context. Each checkpoint represents a complete, reproducible snapshot of the multi-agent system state.
  \item \textbf{Dependency management}: Explicit dependencies in execution plans; parallel execution where independent; sequential where dependencies exist
  \item \textbf{Authority paths}: Decision-making follows a strict hierarchy, never horizontal consensus. Veto authority flows upward: when a critic rejects output, authority returns to the internal team to retry without propagating failure upward or seeking consensus from peers. Approvals flow upward: when a critic approves, the result escalates to the next decision point (parent coordinator, final output gate, or user). This asymmetry ensures that rejections are handled locally within team boundaries while approvals enable work to progress.

  \item \textbf{Escalation}: Failures are contained and resolved at the lowest appropriate level. Persistent internal team failure (repeated rejections despite retries) escalates to the user for replanning, acknowledging that the team cannot solve the problem under the current plan. Policy violations (guardrails triggered, security constraints violated, compliance checks failed) immediately escalate to a user gate for explicit approval before proceeding. Unresolvable contradictions (critics disagree fundamentally, different validators reject for conflicting reasons) escalate to human review, recognizing that the system cannot adjudicate the conflict autonomously.
\end{itemize}

\begin{figure*}
\centering
\includegraphics[width=\linewidth]{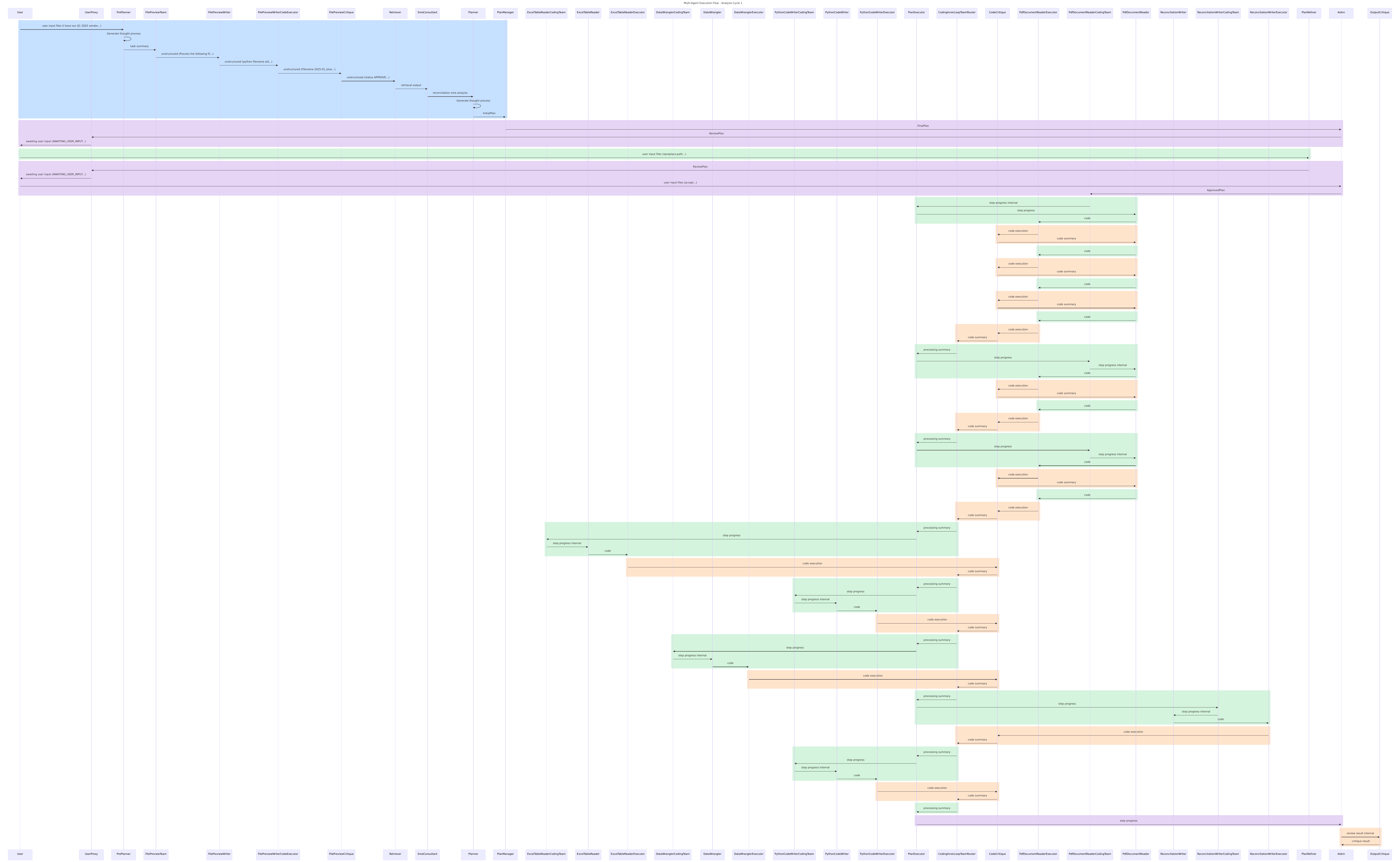}
\caption{Multi-Agent Execution Flow - Analysis Cycle 1. Participants are color-coded by architectural phase: gray (user layer), blue (planning), green (execution), orange (validation), and purple (coordination). Interactions are grouped by phase with colored boxes showing the progression from planning through specialized execution teams to multi-layer validation before user presentation.}
\label{fig:execution-flow}
\end{figure*}

\subsection{Team-Based Execution with Pre-Declared Acceptance Testing}
\label{sec:team-execution}

\subsubsection{Role-Based Execution Structure}

Execution follows a fixed, sequenced pattern: \textbf{Planner first, Critics last, with iterative loops in between}.

\textbf{Stage 1: Planning.} The planner analyzes the user request, identifies required data sources and domain expertise, and constructs a detailed execution DAG. Critically, the planner pre-declares acceptance criteria upfront in the plan. These criteria are not emergent or subjective; they are explicit decision gates that critics will apply. The plan specifies what success looks like before any work is performed.

\textbf{Stage 2: Human approval.} The execution plan is presented to the user for review and explicit approval. No execution proceeds without human sign-off on the plan and its criteria. This is an architecturally enforced gate; the system will not execute without user approval.

\textbf{Stage 3: Isolated execution.} Upon approval, the plan executor orchestrates specialized teams (writers, executors, summarizers) with strict context isolation. Team chatter does not contaminate the parent planning scope. Writers generate code; executors run it; results are aggregated and passed to critics.

\textbf{Stage 4: Critic evaluation and iterative loop.} Critics evaluate outputs against the pre-declared criteria. If approved, results advance. If rejected, authority returns to the internal team to retry without replanning. This loop repeats until critics approve, criteria are met, or persistent failure triggers escalation to the user.

\subsubsection{Veto Authority and Acceptance Criteria}

The architecture applies the Swiss cheese model: multiple validation layers with misaligned failure modes prevent errors from propagating to users.

\begin{itemize}
  \item \textbf{Pre-declared acceptance}: Acceptance criteria are defined upfront in the plan, not emergent or subjective. Critics evaluate against these explicit gates.
  \item \textbf{Critic veto}: The critique agent can reject outputs entirely, triggering internal team retry without re-planning. This catches errors before they propagate downstream.
  \item \textbf{Authority hierarchy}: Code writers cannot declare their own success. Executors cannot declare success. Only independent critics can approve work for advancement.
  \item \textbf{No self-certification}: Passing automated tests is insufficient. Outputs must meet pre-declared criteria evaluated by a critic, providing an imperfect but independent layer that catches errors tests miss.
  \item \textbf{Iterative refinement}: Failed critiques trigger re-execution within team boundaries, while successful critiques propagate results upward. Because critics have different failure modes than producers, errors rarely escape all layers.
  \item \textbf{Escalation}: Persistent rejection after multiple retries escalates to the user for replanning.
\end{itemize}

\subsection{Remote Code Execution: Separating Brains from Hands}
\label{sec:data-kernel}

The remote code executor provides a critical separation: reasoning models (the brains) never directly touch raw data or tool outputs. Instead of agents calling tools directly (which would inject full API responses into their context), agents write code that invokes tools like MCP servers; those tool responses remain in the remote execution layer. Inspired by operating system abstractions for resource management~\cite{Mei2024AIOS}, this architecture maintains clean boundaries between perception and execution, preventing context contamination while enabling scalable data transformations and API interactions.

\subsubsection{Core Responsibility}

Maintains separation between reasoning (perception) and execution (data transformation + tool invocations):

\begin{itemize}
  \item \textbf{Context isolation}: Raw data and tool outputs never enter agent context windows; agents perceive summaries, schemas, and sample rows rather than full datasets or API responses; prevents context contamination
  \item \textbf{Code-based tool invocation}: Agents write code to invoke tools (MCP servers, APIs) rather than calling them directly; tool responses remain in remote execution layer; only relevant extracts return to agent context
  \item \textbf{Remote execution}: Transformation code and tool invocations execute in isolated environment; heavy data manipulation and API calls happen away from reasoning models; separation of brains (agents) from hands (executor)
  \item \textbf{Input standardization}: Connectors convert all sources (Snowflake, CSV, Stripe, MCP) to unified schema; agents reason about consistent abstractions
  \item \textbf{Routing}: The Plan executor routes transformation requests to appropriate teams containing code writers, executors and critiques.
  \item \textbf{Synthesis}: Aggregates results from multiple agents; returns only relevant summaries to reasoning context; conflict resolution; deduplication
  \item \textbf{Integrity}: The summarizers extract column-level lineage for each step, explain the math in LaTeX symbols for complex operations and record information needed to verify references across transformations
  \item \textbf{Optimization}: The agents can optimize operations to reuse intermediate results, recompute new summaries from them or load data back into source systems as temporary tables for further joins or filtered aggregations
\end{itemize}

\subsubsection{Transformation Primitives (50+ Operations)}

Organized as vowels/consonants/punctuation composable alphabet:

\begin{itemize}
  \item \textbf{Vowels} (semantic high-level): Entity extraction, relationship discovery, concept normalization, semantic similarity, temporal alignment
  \item \textbf{Consonants} (structural): Filter, select, project, join, aggregate, group-by, pivot, unpivot, union, difference, distinct, sort, limit, window functions, type conversions
  \item \textbf{Punctuation} (control flow): Conditional execution, branching, loop-until, parallel map, error handling, fallback paths, logging checkpoints, audit trails
  \item \textbf{Composition}: Finite set of primitives (vowels + consonants + punctuation) compose into infinite possible data workflows; enables both simple (one-step) and complex (multi-step) operations
\end{itemize}

\subsubsection{Context Isolation and Optimization}

The remote code executor enforces strict boundaries: execution results return only what the planner explicitly requested, preventing context leakage from the execution layer back to the planning context. When a planner asks for a query result, the executor returns summaries and statistics, not raw data rows. When intermediate steps generate artifacts or side-effects, these remain in the execution context and are not injected back into the reasoning layer unless explicitly extracted via specific requests.

Critically, results are held by the executor and remain available for subsequent planning phases. A planner can request alternative summaries of already-computed results: asking for a category breakdown by State when the executor previously returned aggregated totals, or requesting a distribution function over revenue numbers. This selective access allows planners to explore the execution results through different lenses without leaking the full working dataset into planning context.\reviewnote{Security question: Who has access to data in the remote execution layer? Are there data governance policies? Can one user's executor access another user's data? Needs discussion of multi-tenancy + data isolation beyond what's mentioned.}

This isolation serves two purposes. First, it minimizes the working set of information available to planning agents, reducing hallucination surface and token consumption. Planners reason about data shape and content without holding the full dataset in context. Second, it prevents unintended coupling where execution details (transient errors, intermediate states, implementation choices) contaminate planning decisions. The planner's context remains clean, focused on the high-level task specification.

Future work will quantify the context minimization achieved through this selective integration, measuring context window consumption reduction and its impact on both token efficiency and hallucination rates across execution stages.

\subsection{Comparison: Traditional Single-Agent vs. AI Office}
\label{sec:comparison}

Table~\ref{tab:comparison} summarizes the key differences between traditional single-agent systems and our AI Office architecture across nine dimensions, organized by architectural foundations, execution pipeline, and production readiness.

\begin{table*}[htbp]
\centering
\caption{Comparison of Single-Agent vs. AI Office Architecture}
\label{tab:comparison}
\small
\begin{tabular}{@{}p{2.4cm}p{5.3cm}p{7.8cm}@{}}
\toprule
\textbf{Dimension} & \textbf{Single-Agent} & \textbf{AI Office} \\
\midrule
\multicolumn{3}{l}{\textit{Execution Pipeline}} \\
\addlinespace
Context hygiene & Full conversation history to all components; context contamination compounds errors & Message filtering per agent role: planners see user intent and retrieved context; executors see step plans and prior outputs; critics see execution results. No cross-contamination between phases \\
\addlinespace
Data isolation & Raw data in LLM context (privacy risk, context overflow) & Data never touches LLM; agents receive schemas and summaries only \\
\addlinespace
Data grounding & LLM generates answers directly (hallucination-prone, unverifiable claims) & Agents generate code; Jupyter executes against real data sources. Every number traceable to executed query, not confabulation \\
\addlinespace
Correctness & Self-review insufficient; errors propagate unchecked to output & Multi-layer validation with domain-specialized critics; any layer can veto and trigger retry before propagation \\
\midrule
\multicolumn{3}{l}{\textit{Production Readiness}} \\
\addlinespace
Resilience & Single failure terminates task; no recovery path & Graceful degradation: model upgrade on failure, cross-provider fallback, checkpoint-based resume from any decision point, escalation to human on persistent failure \\
\addlinespace
Extensibility & Monolithic prompt; changes risk regressions across capabilities & FSM-based routing; add agents, writers, or critics without modifying core. 50+ agents, 32 data writers, 15+ sources, each independently testable and evolvable \\
\addlinespace
Auditability & Opaque reasoning; no decision trail & Event-sourced SessionLog with column-level lineage. Work backwards from any result; trace decisions through agent chain; exposure analysis when upstream data changes \\
\bottomrule
\end{tabular}
\end{table*}

\subsubsection{Key Architectural Differentiators}

\textbf{Adaptive model allocation.}\label{sec:adaptive-allocation} Like staffing a team: junior analysts handle routine work; senior specialists engage when complexity demands. Writers start with fast, economical models; upgrade only on critique failure. Critics run on different models than writers for cognitive diversity, not self-review. Cross-provider fallback ensures resilience. Cost and latency scale with task difficulty: cheap and fast by default, thorough when stakes demand it.

\textbf{Separation of perception and execution.} LLMs orchestrate like expert analysts directing work, not clerks processing spreadsheets. Agents write code; Jupyter executes against real data; only schemas and summaries return to context. Raw data never touches LLM, solving context limits (working set $\gg$ context window), data sensitivity (PII stays in execution layer), and hallucination (answers grounded in code execution, not confabulation).

\textbf{Strategic LLM boundaries.} Single-agent systems route everything through LLMs, compounding unreliability. AI Office deploys LLMs surgically for reasoning (planning, critique, summarization) while deterministic code handles orchestration: FSM routing, state reconstruction, step sequencing. PlanExecutor never asks an LLM ``what's next.''

\textbf{Hierarchical veto, not consensus voting.} Specialized critics (CodeCritique for standards, ChartCritique for visualization, OutputCritique for user intent, PlanCritique for plan soundness including dataflow integrity, goal alignment, and domain constraints) each hold independent veto authority in their domain; like auditors who can each halt a release for different violations, unanimous approval is required for advancement, and new domain critics slot in without modifying existing validators.

%% file: sections/evaluation.tex
\section{Use Cases and Evaluation}
\label{sec:evaluation}

\subsection{Financial Domain Case Study}
\label{sec:finance}

\subsubsection{Problem Statement}

Accounts payable reconciliation requires matching vendor invoices against recorded expenses, a task complicated by vendor name variations, split payments across periods, and timing differences between invoice dates and payment records. We evaluate the multi-agent architecture on a representative Q1 2025 reconciliation task: matching 9 PDF invoices from 3 vendors (healthcare, telecommunications, cloud services) against a QuickBooks Online expense summary.

The consequences of extraction errors in this domain are significant. Missing invoice numbers break audit trails and prevent duplicate detection. Incorrect amounts cause financial misstatements that propagate to management reports and regulatory filings. Wrong dates lead to period misallocation, affecting month-end close accuracy. In traditional single-agent pipelines, such errors would require manual review of raw documents, a process that scales poorly when reconciling hundreds of invoices monthly.

The architecture generalizes beyond this representative session: production deployments process hundreds of invoices per cycle, query accounting systems directly via API rather than exported spreadsheets, perform sub-line-item reconciliation matching individual charges to expense categories, and conduct trend analysis identifying month-over-month outliers for anomaly detection.

\subsubsection{System Application}

\begin{figure*}[t]
\centering
\includegraphics[width=\textwidth]{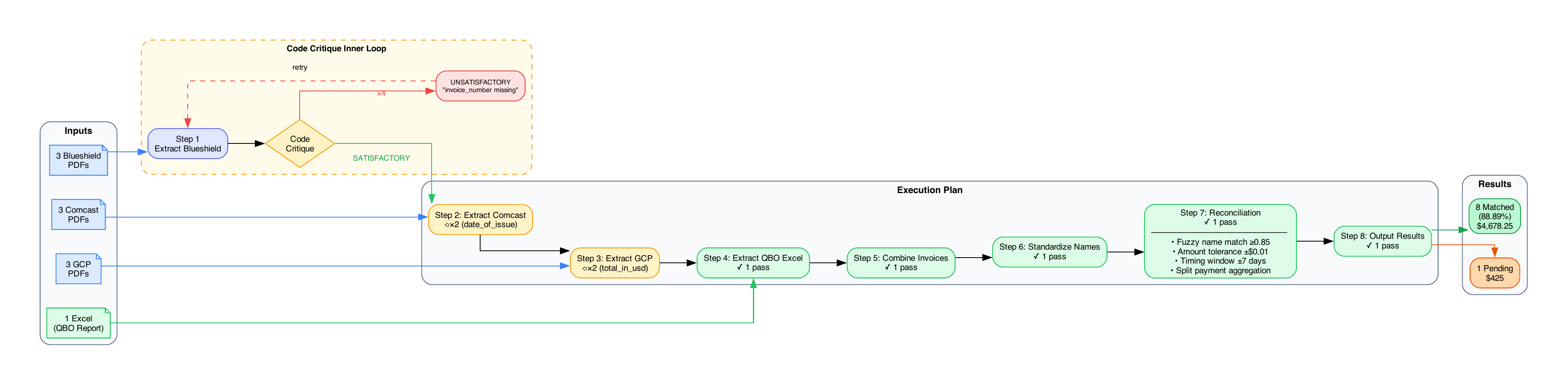}
\caption{Financial reconciliation pipeline execution. The Code Critique inner loop (left, dashed) shows Step 1 requiring 4 iterations to extract missing \texttt{invoice\_number} fields. Steps 2--3 similarly required 2 iterations each. Steps 4--8 passed on first attempt. The reconciliation step applies fuzzy matching ($\geq$0.85), amount tolerance ($\pm$\$0.01), and timing windows ($\pm$7 days) to produce an 88.89\% match rate.}
\label{fig:financial-pipeline}
\end{figure*}

The system decomposed the reconciliation task into an 8-step execution plan (Figure~\ref{fig:financial-pipeline}): three parallel PDF extraction steps (one per vendor schema), Excel parsing for the expense report, invoice consolidation, vendor name standardization with fuzzy matching, reconciliation logic applying configurable thresholds, and output generation. Each step executed within the inner-loop critique cycle described in Section~\ref{sec:metrics}.

The Code Critique layer detected extraction failures that would otherwise propagate silently. During PDF parsing, the critic identified missing required fields: \texttt{invoice\_number} absent from healthcare invoices (preventing audit trail linkage), \texttt{date\_of\_issue} missing from telecommunications invoices (causing temporal misalignment), and \texttt{total\_in\_usd} unparsed from cloud service invoices (yielding null amounts). Each failure triggered automatic retry with refined extraction logic, requiring 4, 2, and 2 iterations respectively before satisfying acceptance criteria.

\subsubsection{Results}

The reconciliation achieved an 88.89\% match rate, successfully matching 8 of 9 invoices totaling \$4,678.25 against recorded expenses. One invoice (\$425) remained unmatched due to a payment pending in the subsequent period (a legitimate business condition rather than system error). The system correctly categorized this as \texttt{PAYMENT\_PENDING} based on recency heuristics.

\begin{table}[H]
\centering
\caption{Extraction Errors Caught by Code Critique}
\label{tab:extraction-errors}
\small
\begin{tabular}{@{}llcl@{}}
\toprule
\textbf{Vendor Type} & \textbf{Missing Field} & \textbf{Iterations} & \textbf{Consequence if Missed} \\
\midrule
Healthcare & \texttt{invoice\_number} & 4 & Broken audit trail \\
Telecom & \texttt{date\_of\_issue} & 2 & Period misallocation \\
Cloud & \texttt{total\_in\_usd} & 2 & Null amount propagation \\
\bottomrule
\end{tabular}
\end{table}

Without multi-agent verification, all three extraction errors would have produced structurally valid but semantically incorrect output: DataFrames with proper schemas but null or missing values in critical fields. These errors are particularly insidious: downstream reconciliation logic would execute without exception, producing plausible but incorrect match rates. The 5 automatic retries (13 total critique iterations across 8 steps) represent errors that would otherwise require manual debugging after observing unexpected reconciliation failures.

Total session time was 12.8 minutes, acceptable for batch reconciliation workflows where correctness outweighs latency. The critique overhead (approximately 40\% of session time on retry iterations) is justified by the alternative: manual review cycles measured in hours when extraction errors surface during financial close.

\subsection{Evaluation Metrics and Analysis}
\label{sec:metrics}

We evaluated the multi-agent architecture across 522 production sessions to measure recovery effectiveness and cost-benefit tradeoffs.

\subsubsection{Probabilistic Analysis of Cascaded Critique Layers}

The multi-agent architecture employs a three-layer cascaded critique system: (1) Code Critique for code generation tasks, (2) Chart Critique for visualization tasks, and (3) Output Critique for holistic validation. We model each layer's effectiveness using conditional probability to quantify their individual and compound contributions.

\begin{table}[H]
\centering
\caption{Three-Layer Critique System Effectiveness}
\label{tab:critique-effectiveness}
\footnotesize
\begin{tabular}{@{}llrrrr@{}}
\toprule
\textbf{Layer} & \textbf{Scope} & \textbf{Input} & \textbf{Catch} & \textbf{Rate} & \textbf{Cumul.} \\
\midrule
L0: First Pass & Clean pass & 522 & 130 & 24.9\% & 24.9\% \\
L1: Code & Code errors & 392 & 337 & 86.0\% & --- \\
L1: Chart & Chart errors & 392 & 7 & 1.8\% & --- \\
L1: Combined & Inner loop & 392 & 344 & 87.8\% & 90.8\% \\
L2: Output & Output quality & 48 & 7 & 14.6\% & 92.1\% \\
Residual & User rejected & 41 & --- & --- & 7.9\% \\
\bottomrule
\end{tabular}
\end{table}

Let $P(E) = 0.75$ denote the base error rate (392/522 sessions required critique intervention).\reviewnote{Good: directly show the base failure rate. This justifies the multi-layer approach. Consider breaking down what types of errors comprise this 75\%---code bugs, logic errors, data mismatches, etc.} For the inner loop (L1), Code and Chart critiques address distinct failure modes: Code Critique catches 337 sessions (86.0\%) involving syntax errors, logic bugs, and API misuse, while Chart Critique catches 7 sessions (1.8\%) with visualization-specific issues. The combined inner loop catch rate is $P(\text{catch}_1) = 344/392 = 87.8\%$.

The Output Critique (L2) operates on the 48 sessions that escaped the inner loop, achieving $P(\text{catch}_2) = 7/48 = 14.6\%$. The probability that an error escapes all layers is:

\begin{align}
P(\text{user rej.}) &= P(E) \times (1 - P(\text{catch}_1)) \times (1 - P(\text{catch}_2)) \notag \\
&= 0.75 \times 0.122 \times 0.854 = 0.078
\end{align}

This matches our observed 7.9\% user disapproval rate (41/522). The cascaded probability model assumes conditional independence between critique layers; that is, an error's probability of escaping L2 is independent of whether it escaped L1. While the empirical specialization of critics (Code vs. Chart vs. Output) suggests low overlap in failure modes, this independence assumption is an approximation used for tractability rather than a proven property.

\textbf{Diminishing Returns Analysis.}\reviewnote{Strong analysis showing optimality. This is a key contribution---you've identified the practical ceiling for automated verification. Make this even more prominent in the abstract/conclusion.} Each successive layer catches a smaller absolute number of errors: the Inner Loop (L1) saves 344 sessions (+65.9\%). Code Critique catches 337 sessions (86.0\% of errors), while Chart Critique catches 7 sessions (1.8\%) addressing visualization-specific issues such as incorrect chart types and axis configurations. Output Critique (L2) adds 7 saves (+1.3\%). A hypothetical fourth layer with equivalent effectiveness to L2 would reduce the escape rate from 7.9\% to 6.7\%, saving only 6 additional sessions while incurring additional latency and compute costs. More critically, errors that escape all three critics exhibit qualitatively different characteristics: requirement ambiguity (correct implementation of misunderstood intent), subjective preferences (technically valid but stylistically misaligned), and domain-specific edge cases. These error types are fundamentally resistant to automated critique and require human-in-the-loop resolution. The observed 7.9\% residual thus represents a practical irreducible floor for automated systems, making three critics across two layers an optimal balance between quality assurance and system efficiency.

\subsubsection{Session Recovery Flow}

\begin{figure}[t]
\centering
\includegraphics[width=1.15\linewidth]{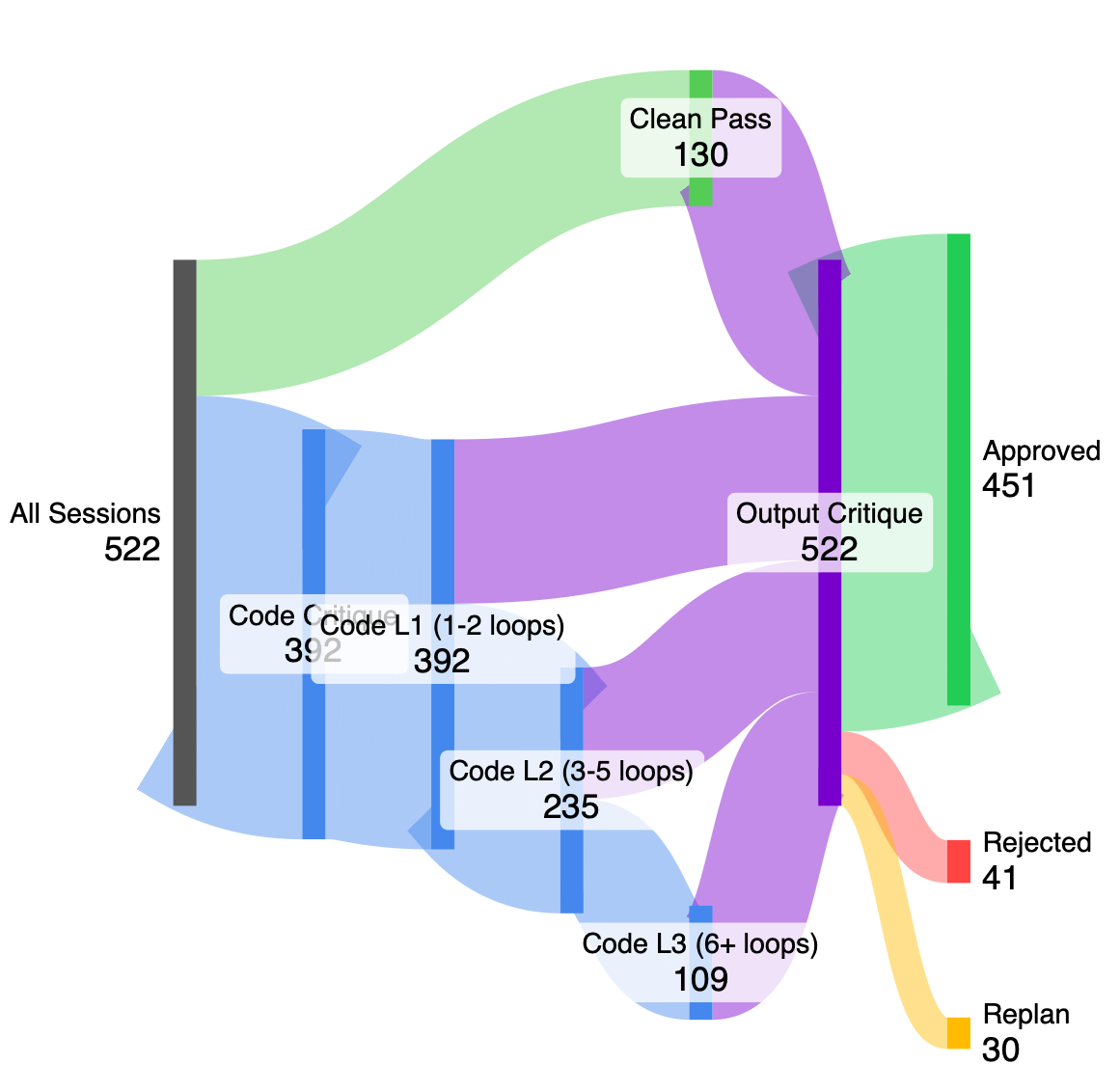}
\caption{Multi-agent session recovery flow. Of 522 sessions, 130 succeeded on first pass, 344 recovered through Code Critique (337) and Chart Critique (7), 7 recovered via Output Critique, and 41 were rejected by users.}
\label{fig:session-recovery}
\end{figure}

Figure~\ref{fig:session-recovery} illustrates the session recovery flow across 522 production sessions.

The system achieved a 92.1\% overall success rate. Of sessions requiring recovery, 40.1\% resolved within 1--2 additional iterations (quick recovery), demonstrating the efficiency of the critic-triggered retry mechanism. Only 7.9\% of sessions ended in user disapproval, indicating that internal quality gates successfully filter most errors before user exposure. An additional 30 sessions (5.7\%) triggered a replan request, where the system identified missing information and prompted users for additional inputs (e.g., supplementary documents, clarifying parameters) before proceeding with analysis.

\subsubsection{Cost-Benefit Analysis}

Tables~\ref{tab:recovery-credits} and~\ref{tab:recovery-time} quantify recovery cost across iteration levels for the 392 sessions that required retry.

\begin{table}[H]
\centering
\caption{Credit Cost by Recovery Level (n=392 sessions)}
\label{tab:recovery-credits}
\small
\begin{tabular}{@{}lrrrrr@{}}
\toprule
\textbf{Recovery Level} & \textbf{Sessions} & \textbf{Total} & \textbf{Recovery} & \textbf{\%} \\
\midrule
Level 1 (1--2 extra) & 157 & 3,504.5 & 686.8 & 19.6 \\
Level 2 (3--5 extra) & 126 & 5,756.4 & 1,831.2 & 31.8 \\
Level 3 (6+ extra) & 109 & 11,065.0 & 5,324.1 & 48.1 \\
\midrule
\textbf{Total} & \textbf{392} & \textbf{20,325.9} & \textbf{7,842.1} & \textbf{38.6} \\
\bottomrule
\end{tabular}
\par\vspace{2pt}\noindent\footnotesize{Credits charged to customer. Recovery = credits on iterations $>$ 1.}
\end{table}

\begin{table}[H]
\centering
\caption{Time Cost by Recovery Level (n=392 sessions)}
\label{tab:recovery-time}
\small
\begin{tabular}{@{}lrrrrr@{}}
\toprule
\textbf{Recovery Level} & \textbf{Sessions} & \textbf{Total (hrs)} & \textbf{Recovery} & \textbf{\%} \\
\midrule
Level 1 (1--2 extra) & 157 & 12.3 & 1.2 & 10.0 \\
Level 2 (3--5 extra) & 126 & 14.4 & 3.1 & 21.6 \\
Level 3 (6+ extra) & 109 & 26.0 & 7.1 & 27.4 \\
\midrule
\textbf{Total} & \textbf{392} & \textbf{52.7} & \textbf{11.5} & \textbf{21.8} \\
\bottomrule
\end{tabular}
\par\vspace{2pt}\noindent\footnotesize{Time = active processing time (excludes user idle gaps $>$30s).}
\end{table}

Key findings from the recovery analysis:

\begin{itemize}
  \item \textbf{40\% of sessions} (157/392) recovered within 1--2 extra iterations with only 19.6\% credit cost and 10.0\% time cost, demonstrating efficient error correction for most cases.
  \item \textbf{Token cost exceeds time cost}: Recovery consumes 38.6\% of credits but only 21.8\% of active time, as LLM calls dominate cost while execution is fast.
  \item \textbf{Heavy-tail distribution}: Level 3 sessions (6+ extra iterations) represent 28\% of sessions (109/392) and consume 68\% of recovery credits but only 62\% of recovery time.
  \item \textbf{Total recovery investment}: 7,842 credits and 11.5 hours of active processing to achieve 92.1\% success rate across 392 sessions that required code critique.
\end{itemize}

\begin{figure}[t]
\centering
\begin{tikzpicture}
\begin{axis}[
    ybar stacked,
    width=0.9\linewidth,
    height=4.5cm,
    ylabel={Credits},
    symbolic x coords={Level 1\\(1-2 extra), Level 2\\(3-5 extra), Level 3\\(6+ extra)},
    xtick=data,
    x tick label style={align=center, font=\small},
    ymin=0,
    ymax=13000,
    legend style={at={(0.5,-0.28)}, anchor=north, legend columns=2, font=\small},
    bar width=24pt,
    enlarge x limits=0.3,
]
\addplot[fill=blue!70, draw=blue!80] coordinates {
    ({Level 1\\(1-2 extra)}, 2817.7)
    ({Level 2\\(3-5 extra)}, 3925.2)
    ({Level 3\\(6+ extra)}, 5740.9)
};
\addplot[fill=red!70, draw=red!80] coordinates {
    ({Level 1\\(1-2 extra)}, 686.8)
    ({Level 2\\(3-5 extra)}, 1831.2)
    ({Level 3\\(6+ extra)}, 5324.1)
};
\legend{Base Credits, Recovery Credits}
\end{axis}
\end{tikzpicture}
\caption{Credit cost by recovery level. Level 3 sessions account for 68\% of all recovery credits despite representing only 28\% of sessions.}
\label{fig:credit-cost}
\end{figure}
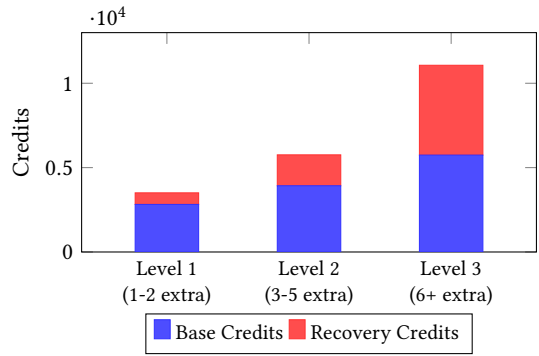

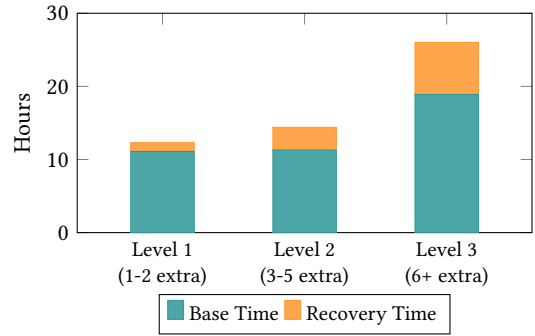
\begin{figure}[t]
\centering
\begin{tikzpicture}
\begin{axis}[
    ybar stacked,
    width=0.9\linewidth,
    height=4.5cm,
    ylabel={Hours},
    symbolic x coords={Level 1\\(1-2 extra), Level 2\\(3-5 extra), Level 3\\(6+ extra)},
    xtick=data,
    x tick label style={align=center, font=\small},
    ymin=0,
    ymax=30,
    legend style={at={(0.5,-0.28)}, anchor=north, legend columns=2, font=\small},
    bar width=24pt,
    enlarge x limits=0.3,
]
\addplot[fill=teal!70, draw=teal!80] coordinates {
    ({Level 1\\(1-2 extra)}, 11.1)
    ({Level 2\\(3-5 extra)}, 11.3)
    ({Level 3\\(6+ extra)}, 18.9)
};
\addplot[fill=orange!70, draw=orange!80] coordinates {
    ({Level 1\\(1-2 extra)}, 1.2)
    ({Level 2\\(3-5 extra)}, 3.1)
    ({Level 3\\(6+ extra)}, 7.1)
};
\legend{Base Time, Recovery Time}
\end{axis}
\end{tikzpicture}
\caption{Time cost by recovery level (active processing time). Level 3 sessions spend 27\% of time in recovery, far less than their 48\% credit share.}
\label{fig:time-cost}
\end{figure}

Figures~\ref{fig:credit-cost} and~\ref{fig:time-cost} visualize the cost breakdown by recovery level, illustrating the disproportionate resource consumption at Level 3.

\subsubsection{Error Detection Rate}

\textbf{Errors Detected Through Consensus.} The cascaded critique system detected and resolved errors in 92.1\% of sessions (481/522). Of sessions requiring intervention, the inner loop (Code and Chart Critique) achieved consensus on 344 sessions (87.8\% of 392), while Output Critique resolved an additional 7 sessions. The remaining 41 sessions (7.9\%) required user intervention to identify issues that automated critics missed.

\textbf{False Negative Rate.} We define false negatives as errors that escape all critique layers and reach user review. The observed false negative rate is 7.9\% (41/522 sessions). These errors exhibit characteristics resistant to automated detection: requirement ambiguity where implementation was technically correct but misaligned with unstated intent, subjective preferences where output quality was acceptable but stylistically mismatched, and domain-specific edge cases requiring context unavailable to critics. This 7.9\% represents a practical floor for the current architecture.

\textbf{False Positive Rate.} False positives (valid outputs wrongly rejected by critics, causing unnecessary retries) cannot be directly measured in production without ground-truth labels for each iteration. Measuring false positives would require determining whether original code was correct before critic-requested modifications, information unavailable post-hoc. However, we note that the iterative retry mechanism is self-correcting: if a critic incorrectly flags valid work, subsequent iterations converge toward approval. Platform-level prompt refinements address systematic false positive patterns as they emerge through operational monitoring. The absence of explicit false positive measurement is a limitation; future work could employ held-out labeled datasets to quantify this rate.

\textbf{Disagreement Patterns.} Code and Chart critiques address orthogonal failure modes with minimal overlap. Of 392 sessions requiring inner-loop intervention, 385 involved Code Critique (syntax, logic, API errors) while 7 involved Chart Critique (visualization issues). This specialization validates the Swiss cheese model: critics with distinct failure modes catch errors that homogeneous review would miss. Cross-critic disagreement (one approving while another rejects) was not observed in this dataset, as critics evaluate different output dimensions rather than the same artifact.

\ifdefined\isrelease
\else
  \subsubsection{Accuracy and Correctness}
  \reviewnote{IMPORTANT: This section must be filled in. Show: (1) Baseline single-agent accuracy on 522 sessions (need counterfactual), (2) Accuracy breakdown by model provider, (3) Accuracy by task complexity. Without this, you cannot claim "92.1\% improvement."}
  {[}TODO: Present:{]}
  \begin{itemize}
    \item Percentage of correct outputs: single-agent vs. consensus
    \item Breakdown by model
    \item Breakdown by task type
    \item Comparative baselines
  \end{itemize}
\fi

\subsection{Comparative Analysis}
\label{sec:comparative}

To contextualize the multi-agent architecture's 92.1\% success rate, we compare against single-agent baselines using the financial reconciliation task from Section 4.1.

\subsubsection{Single-Agent Baseline}

We executed the identical reconciliation task (9 invoices against QBO expense report, \$40 ground-truth discrepancy) using a single-agent approach with a frontier model from a leading vendor.\footnote{Session transcripts available at \url{https://tinyurl.com/yw6dtztn}} Across 10 independent trials with identical inputs:

\begin{itemize}
  \item \textbf{6 trials (60\%)}: Correctly identified the \$40 discrepancy
  \item \textbf{3 trials (30\%)}: Reported incorrect discrepancy amounts
  \item \textbf{1 trial (10\%)}: Reported ``no discrepancy found''
\end{itemize}

Critically, the model asserted high confidence in all 10 trials, providing no signal to distinguish correct from incorrect results. Average completion time was 1--2 minutes.

\subsubsection{Self-Verification Baseline}

We tested whether prompting the model to verify its own output improves accuracy. After initial reconciliation, we prompted: ``please verify your reconciliation.'' Across 5 trials where the model initially identified the correct \$40 discrepancy:

\begin{itemize}
  \item \textbf{3 trials}: Changed from correct to incorrect (``I misinterpreted... no discrepancy'')
  \item \textbf{2 trials}: Maintained the correct answer
\end{itemize}

Self-verification reduced accuracy rather than improving it. The same reasoning process that produced the initial answer cannot reliably evaluate it: the model second-guesses correct conclusions while maintaining confidence in incorrect ones.

These baselines are illustrative rather than exhaustive and are intended to contextualize error modes rather than establish state-of-the-art comparisons.

\subsubsection{Multi-Agent Architecture}

We executed the same financial reconciliation task using our multi-agent architecture across 20 independent trials. The system achieved 90\% accuracy (18/20 correct), with extraction errors automatically detected and corrected through the Code Critique layer.

\begin{table}[H]
\centering
\caption{Single-Agent vs. Multi-Agent Comparison}
\label{tab:comparative}
\footnotesize
\begin{tabular}{@{}lcccc@{}}
\toprule
\textbf{Approach} & \textbf{Trials} & \textbf{Acc.} & \textbf{Time} & \textbf{Error Signal} \\
\midrule
Single-agent & 10 & 60\% & 1--2 min & None \\
Single + self-verify & 5 & $<$60\% & 2--3 min & Self-doubt \\
Multi-agent (ours) & 20 & 90\% & 4.2 min & Auto-retry \\
\bottomrule
\end{tabular}
\end{table}

The multi-agent architecture achieves significantly higher accuracy by introducing external verification through independent critics. Unlike self-verification, the Code Critique agent evaluates outputs using different reasoning than the agent that produced them, enabling detection of errors that self-checking misses. The 2--3$\times$ latency overhead (4.2 min vs. 1--2 min) is justified for workflows where correctness outweighs speed. A \$40 reconciliation error in financial close has consequences far exceeding the cost of additional compute time.

%% file: sections/discussion.tex
\section{Discussion}
\label{sec:discussion}

\subsection{Key Insights}

\subsubsection{Orthogonal Failure Modes Validate Role Specialization}

The inner loop's 87.8\% catch rate reflects a structural property: Code Critique and Chart Critique address fundamentally different failures. Of 392 sessions requiring inner-loop intervention, 385 involved Code Critique while only 7 involved Chart Critique (1.8\%), and these populations are mutually exclusive. This independence validates the Swiss cheese model empirically: specialists with different focus areas catch errors that homogeneous review would miss.

The implication for critic design: prioritize orthogonality over redundancy. A fourth critic duplicating Code Critique's failure modes catches few additional errors. A critic targeting a genuinely different class (data lineage validation, regulatory compliance) extends coverage into unprotected territory.

\subsubsection{The 7\% Residual: Automation's Ceiling}

The 41 sessions where all critics approved but users rejected share a common characteristic: they require information absent from the execution context.

\begin{itemize}
  \item \textbf{Requirement ambiguity}: Correct implementation of incorrectly stated intent
  \item \textbf{Subjective preferences}: Technically valid but stylistically misaligned (chart colors, verbosity)
  \item \textbf{Domain edge cases}: Correct methodology misapplied to unusual patterns (fiscal year boundaries, industry-specific treatments)
\end{itemize}

A hypothetical fourth layer with equivalent effectiveness to Output Critique (14.6\%) would reduce the escape rate from 7.9\% to 6.7\%, saving only 6 additional sessions at significant cost. The 7\% floor represents errors that automated systems cannot resolve without external input. This suggests 93\% as the practical ceiling for automated verification on tasks of this complexity.

\subsubsection{When Overhead is Justified}

The roughly 40\% credit overhead buys error \textit{containment}: mistakes die in committee rather than reaching users. For financial analysis, an incorrect calculation driving poor decisions costs far more than compute overhead. The cost-benefit calculus depends on error severity. High-stakes tasks (regulatory filings, board presentations) justify full validation; low-stakes tasks (exploratory analysis) may not.

\subsection{Limitations}

\textbf{Domain specificity.} Our evaluation covers financial analysis exclusively. The architectural principles should transfer, but catch rates may differ in domains with more subjective outputs or less well-defined correctness criteria.

\textbf{Cost at scale.} This overhead becomes significant at high volume. Production deployments must route high-stakes work through full validation while allowing routine tasks lighter-weight paths.

\textbf{Latency.} Each critic layer adds sequential latency. For real-time or interactive use cases, this tradeoff may be unacceptable without parallel critique execution.

\textbf{The 7\% floor.} For applications requiring near-perfect accuracy (drug discovery, safety-critical systems), 93\% success is insufficient. These domains require human review of all outputs or fundamentally different verification approaches.

\subsection{Implications for Production AI}

\subsubsection{Deployment Decisions}

Three factors determine suitability:

\begin{enumerate}
  \item \textbf{Error tolerance}: If the residual 7\% error rate is acceptable, deploy with confidence. If sub-1\% is required, plan for human-in-the-loop review of all outputs.
  \item \textbf{Cost sensitivity}: The cost premium is justified when error costs exceed compute costs. This is typically true for high-value, low-volume tasks.
  \item \textbf{Latency requirements}: Multi-layer critique adds sequential delay. Real-time applications need parallel critique or reduced coverage.
\end{enumerate}

\subsubsection{Operational Considerations}

The stage-gated architecture provides natural observability: each critic decision, retry, and escalation generates structured events enabling quality dashboards, cost attribution, and failure analysis. The event-sourced SessionLog supports graceful degradation. Model fallback, checkpoint recovery, and escalation paths ensure partial failures degrade to slower operation rather than incorrect output.

\subsection{Future Work}

\textbf{Adaptive critic selection.} Learning which critic configurations optimize for different task types could reduce overhead while maintaining catch rates. Sessions with complex joins might benefit from data validation critics; visualization-heavy tasks might warrant expanded chart critique.

\textbf{Cross-domain evaluation.} Validating across legal document analysis, medical record summarization, code generation, and general data analysis tasks would establish generalizability bounds and inform domain-specific critic design.

\textbf{Parallel critique.} For independent critics, concurrent execution could reduce latency without sacrificing coverage.

\textbf{Formal failure mode analysis.} Formalizing the independence between critic failure modes, whether through information-theoretic or causal analysis, would provide theoretical grounding for principled critic design beyond empirical trial-and-error.

%% file: sections/conclusion.tex
\section{Conclusion}
\label{sec:conclusion}

We demonstrate that organizational principles of reliability transfer directly to AI system design. Coherence emerges not from a single optimized agent, but from opposing forces holding outputs within acceptable boundaries. In production systems, a Planner pushes for clarity, an Executor for pragmatic implementation, and Critics for correctness. These conflicting incentives create consensus boundaries that prevent drift.

Evaluated on 522 production financial analysis sessions, our multi-agent architecture achieves \textbf{92.1\% success rate}, reducing the baseline error rate from 75\% to 7.9\% residual. This 67 percentage-point improvement comes from cascaded critique: an inner loop (Code and Chart critics) catches 87.8\% of errors through orthogonal specialization, while an outer loop (Output Critique) saves an additional 14.6\% of remaining failures. Single-agent approaches achieve only 60\% accuracy on the same tasks, with no reliable error signal. Self-verification degrades accuracy further; the same reasoning process that produced an answer cannot reliably evaluate it.

The near-independence of Code Critique (86.0\% catch rate on syntax, logic, and API misuse) and Chart Critique (1.8\% on visualization issues) validates the Swiss cheese model empirically. Specialists with different failure modes catch errors that homogeneous review misses. This orthogonality has practical implications: a fourth critic layer with equivalent effectiveness would save only 6 additional sessions (reducing escape rate from 7.9\% to 6.7\%) while adding latency and cost. The 7.9\% residual represents errors fundamentally resistant to automated critique: requirement ambiguity, subjective preferences, and domain edge cases that require external context unavailable to any critic.

This reliability costs 38.6\% computational overhead, measured across 392 sessions requiring recovery. Token costs (38.6\% of credits) exceed time costs (21.8\% of active hours), indicating LLM calls dominate expense. The distribution is heavy-tailed: 40\% of recovery sessions resolve within 1--2 iterations at 19.6\% credit cost, while the most complex 28\% consume 68\% of recovery credits. For high-stakes financial tasks where a \$40 reconciliation error propagates to regulatory filings, this overhead is justified. For exploratory analysis, lighter-weight paths may be appropriate. Production deployments can route requests strategically: full validation for board presentations and regulatory filings, lighter paths for exploratory queries where iteration cost exceeds error cost.

The multi-agent architecture offers an underappreciated benefit: it sidesteps the complexity trap of monolithic prompts. A single-agent system serving diverse tasks accumulates instructions, edge cases, and domain knowledge into an ever-growing system prompt. This bloat degrades performance, increases latency, and makes maintenance fragile. By contrast, each agent in our architecture carries a focused prompt for its specific role. The Planner knows planning; the Code Critique knows code review; the Chart Critique knows visualization standards. Adding capability means adding a new specialist agent, not appending paragraphs to a central prompt. The system grows by composition rather than accumulation.

This composability extends to model selection. Different agents can use different providers, matching model strengths to task requirements. A planning agent benefits from strong reasoning; a code writer benefits from training on code corpora; a summarizer benefits from concision. When a new model releases with improved capabilities in a specific domain, we swap one agent's backend without touching the others. The architecture treats models as interchangeable components rather than monolithic dependencies. This multi-vendor strategy provides resilience against provider outages, pricing changes, and capability regressions while enabling continuous improvement as the model landscape evolves.

The deployment decision reduces to a single question: \textit{Is your domain tolerant of 7.9\% residual error?} For mission-critical systems where error consequences are severe (financial close, regulatory reporting, operational decisions), the answer is typically yes, and the architecture supports confident deployment. For domains requiring near-perfect accuracy (drug discovery, safety-critical control systems), 92.1\% is insufficient and human review remains necessary.

Our results are grounded in financial analysis. The architectural principles (role specialization, orthogonal critique, stage-gated oversight) should transfer to domains with well-defined correctness criteria: legal document analysis, medical record extraction, code generation. Catch rates will differ with output subjectivity. Tasks with fuzzy success criteria (creative writing, open-ended research) may not benefit equally. Cross-domain evaluation remains essential.

Three limitations constrain interpretation. First, we evaluated 522 sessions from a single domain; generalization across problem types is plausible but unproven. Second, 40\% overhead becomes prohibitive at high volume; production systems must route requests strategically. Third, sequential critique adds latency (e.g., 12.8 minutes for complex financial reconciliation) unsuitable for real-time applications without parallel execution.

Beyond error reduction, the architecture provides secondary benefits that practitioners value: natural observability (each critic decision generates structured audit events), graceful degradation (checkpointing enables recovery from partial failures), and data isolation (raw data never enters agent context, simplifying multi-tenancy and compliance).

Open questions remain. Can adaptive critic selection optimize for task characteristics? What is the false positive rate; how many valid outputs did critics wrongly reject? Can parallel execution reduce latency without sacrificing coverage? Does formal failure-mode analysis provide principled guidance for critic design beyond empirical trial?

This work demonstrates that organizational reliability principles translate into concrete AI architectures. The Swiss cheese model, veto authority, and stage-gated oversight are not metaphors but mechanisms. As LLM capabilities improve, the limiting factor shifts from raw capability to coherence and verification. Multi-agent architectures that orchestrate teams of rivals, each with veto authority over acceptable outputs, provide a practical path to production reliability.

%% file: main.bib
@inproceedings{Wang2024MultiPersona,
  title={Unleashing the emergent cognitive synergy in large language models: A task-solving agent through multi-persona self-collaboration},
  author={Wang, Zhenhailong and Mao, Shaoguang and Wu, Wenshan and Ge, Tao and Wei, Furu and Ji, Heng},
  booktitle={Proceedings of the 2024 Conference of the North American Chapter of the Association for Computational Linguistics: Human Language Technologies (Volume 1: Long Papers)},
  pages={257--279},
  year={2024}
}

@article{Naik2024,
  title={Probabilistic Consensus through Ensemble Validation: A Framework for LLM Reliability},
  author={Naik, Ninad},
  year={2024},
  eprint={2411.06535},
  archivePrefix={arXiv},
  primaryClass={cs.AI}
}

@article{Cemri2025,
  title={Why Do Multi-Agent LLM Systems Fail?},
  author={Cemri, Mert and Pan, Melissa Z. and Yang, Shuyi and Agrawal, Lakshya A. and Chopra, Bhavya and Tiwari, Rishabh and Keutzer, Kurt and Parameswaran, Aditya and Klein, Dan and Ramchandran, Kannan and Zaharia, Matei and Gonzalez, Joseph E. and Stoica, Ion},
  journal={arXiv preprint arXiv:2503.13657},
  year={2025},
  eprint={2503.13657},
  archivePrefix={arXiv},
  primaryClass={cs.AI}
}

@article{Huang2024Resilience,
  title={On the Resilience of LLM-Based Multi-Agent Collaboration with Faulty Agents},
  author={Huang, Jen-tse and Zhou, Jiaxu and Jin, Tailin and Zhou, Xuhui and Chen, Zixi and Wang, Wenxuan and Yuan, Youliang and Lyu, Michael R. and Sap, Maarten},
  year={2024},
  eprint={2408.00989},
  archivePrefix={arXiv},
  primaryClass={cs.AI}
}

@article{Kamen2025,
  title={Majority Rules: LLM Ensemble is a Winning Approach for Content Categorization},
  author={Kamen, Ariel and Kamen, Yakov},
  year={2025},
  eprint={2511.15714},
  archivePrefix={arXiv},
  primaryClass={cs.AI}
}

@article{Chen2025LLMEnsemble,
  title={Harnessing Multiple Large Language Models: A Survey on LLM Ensemble},
  author={Chen, Zhijun and Li, Jingzheng and Chen, Pengpeng and Li, Zhuoran and Sun, Kai and Luo, Yuankai and Mao, Qianren and Li, Ming and Xiao, Likang and Yang, Dingqi and Ban, Yikun and Sun, Hailong and Yu, Philip S.},
  year={2025},
  eprint={2502.18036},
  archivePrefix={arXiv},
  primaryClass={cs.CL}
}

@article{Sanchez2025,
  title={Multi-LLM Collaboration for Medication Recommendation},
  author={Sanchez, Huascar and Hitaj, Briland and Bergmann, Jules and Briesemeister, Linda},
  year={2025},
  eprint={2512.05066},
  archivePrefix={arXiv},
  primaryClass={cs.LG}
}

@article{Zhang2024AgentSafetyBench,
  title={Agent-SafetyBench: Evaluating the Safety of LLM Agents},
  author={Zhang, Zhexin and Cui, Shiyao and Lu, Yida and Zhou, Jingzhuo and Yang, Junxiao and Wang, Hongning and Huang, Minlie},
  year={2024},
  eprint={2412.14470},
  archivePrefix={arXiv},
  primaryClass={cs.CL}
}

@article{Mohammadi2025,
  title={Evaluation and Benchmarking of LLM Agents: A Survey},
  author={Mohammadi, Mahmoud and Li, Yipeng and Lo, Jane and Yip, Wendy},
  year={2025},
  eprint={2507.21504},
  archivePrefix={arXiv},
  primaryClass={cs.LG}
}

@article{Raza2025TRiSM,
  title={TRiSM for Agentic AI: A Review of Trust, Risk, and Security Management in LLM-based Agentic Multi-Agent Systems},
  author={Raza, Shaina and Sapkota, Ranjan and Karkee, Manoj and Emmanouilidis, Christos},
  year={2025},
  eprint={2506.04133},
  archivePrefix={arXiv},
  primaryClass={cs.AI},
  doi={10.48550/arXiv.2506.04133}
}

@article{Mei2024AIOS,
  title={AIOS: LLM Agent Operating System},
  author={Mei, Kai and Zhu, Xi and Xu, Wujiang and Hua, Wenyue and Jin, Mingyu and Li, Zelong and Xu, Shuyuan and Ye, Ruosong and Ge, Yingqiang and Zhang, Yongfeng},
  year={2024},
  eprint={2403.16971},
  archivePrefix={arXiv},
  primaryClass={cs.OS}
}

@article{Reason1990,
  title={Human error: models and management},
  author={Reason, James},
  journal={BMJ},
  volume={320},
  number={7237},
  pages={768--770},
  year={2000},
  publisher={British Medical Journal Publishing Group}
}

@article{Shannon1948,
  title={A mathematical theory of communication},
  author={Shannon, Claude E.},
  journal={The Bell System Technical Journal},
  volume={27},
  number={3},
  pages={379--423},
  year={1948},
  publisher={Nokia Bell Labs}
}

@inproceedings{Yao2023ReAct,
  title={ReAct: Synergizing Reasoning and Acting in Language Models},
  author={Yao, Shunyu and Zhao, Jeffrey and Yu, Dian and Du, Nan and Shafran, Izhak and Narasimhan, Karthik and Cao, Yuan},
  booktitle={International Conference on Learning Representations},
  year={2023}
}

@misc{Anthropic2025,
  title={Building effective agents},
  author={{Anthropic}},
  year={2024},
  howpublished={Anthropic Research Blog},
  url={https://www.anthropic.com/research/building-effective-agents}
}

@misc{Cognition2025,
  title={Don't Build Multi-Agents},
  author={{Cognition AI}},
  year={2025},
  howpublished={Cognition Blog}
}
